\documentclass[12pt]{iopart}
\usepackage{siunitx}
\usepackage{graphicx}
\begin{document}

\title{Operating Gravitational Wave Detectors far from equilibrium}

\author{Igor Neri$^{1}$\footnote{To whom correspondence should be addressed.},  Miquel L\'opez-Su\'arez$^1$ and Luca Gammaitoni$^1$}

\address{$^1$ NiPS Laboratory, Dipartimento di Fisica e Geologia, Universit\`a degli Studi di Perugia, 06123 Perugia, Italy}
\ead{igor.neri@nipslab.org}

\begin{abstract}
Tiny vibrations of mechanical structures are the main limiting cause in a number of high sensitivity measurement apparatus, chief among them the most sensitive displacement apparatus on earth: gravitational wave interferometers. Such devices are usually operated at equilibrium and small fluctuations are perceived as noise that sets a lower limit to the detection capabilities. An example is the so-called \emph{thermal noise}, ubiquitous and unavoidable. In this letter we present an approach aimed at operating the interferometer out of equilibrium. We show that selective cooling of single modes of the mechanical structure is able to positively impact the measurement sensitivity, in selected frequency ranges. Experiments conducted on thin silica membranes show promising results for the implementation of such technique in next generation gravitational wave detectors.

\end{abstract}

\maketitle

\section{Introduction}
The internal energy of a macroscopic apparatus at thermal equilibrium is shared among all its degrees of freedom each carrying on average $k_BT$ of energy, where $T$ is the equilibrium temperature and $k_B$ is the Boltzman constant. This is true also for such modes as the macroscopic oscillations of springs, pendula, needles, etc. For a given observable such a thermal energy manifests itself as a random fluctuation and is experimentally perceived as a noise affecting its measured value. The quantitative content of this observation, valid in the classical domain, has been summarized by Boltzman in 1876 in the so called Equipartition Theorem. The spectral shape of the thermal fluctuation is one of the main concerns of those interested in building high sensitivity experimental apparatus. Among these a special place is taken by the most sensitive displacement measuring device ever invented: gravitational wave interferometers\cite{AdL,AdV,Geo} that have been recently employed in the first detection of gravitational wave signals from coalescing black holes\cite{1,2} and binary neutron stars\cite{bns}. In addition to thermal noise, a number of different noise sources, usually modelled at equilibrium condition, affect the functioning of the interferometer,  limiting the detector sensitivity in the frequency range between few Hz and few kHz \cite{AdL,AdV,Geo}.
Most often, the spectral shape of such equilibrium fluctuations is determined by the dissipative properties of the dynamics involved. This result, initially obtained by Callen and Welton for the thermal noise in the linear regime\cite{Callen}, represents the content of the Fluctuation-Dissipation Theorem (FDT). Thus, as prescribed by Boltzmann, the total thermal energy of a single mode stays constant and its spectral shape can be modelled by acting on the dissipation function that represents the energy losses. A simple strategy consists in increasing the quality factor so that most of the vibrational energy is confined around the resonance and decreases in regions far from it. Increasing the quality factor is usually a very demanding task that requires a proper selection of low losses materials\cite{4,5,6} and ad-hoc geometries\cite{7,8}.

In this work we propose a novel approach to enhance the sensitivity of an interferometer operating it far from equilibrium. The operation mode consists in two distinct phases: the cooling phase and the recovery phase. In the cooling phase we dynamically cool one resonant mode by applying a properly designed external feedback. Afterwards, during the recovery phase, we show that the signal-to-noise ratio (SNR), in the frequency range where the cooling has been applied, improves.

Active cooling strategies have been proposed earlier as a potential technique to improve detector sensitivity. We notice here that, as observed by Harris and co-workers\cite{Harris}, as long as the detector dynamics is linear, there exists a real-time estimation strategy that reproduces the same measurement record as any arbitrary feedback protocol. In this case any active cooling, stationary or not, does not improve sensitivity over properly chosen data analysis. However, it is important to remember that this is only true when we are dealing with linear oscillators\cite{Harris}. In any real physical system with many degrees of freedom this condition is only approximately met and non-linearities that accounts for energy transfers between different spectral regions are always present. Moreover, even in the linear approximation condition, it may result operatively very difficult to implement the data processing technique, because it requires the knowledge of the transfer function involved. At difference with the data processing technique, our approach can be implemented also without a prior knowledge of the transfer function and in cases where the dynamics is markedly non-linear. The presented approach however does not prevent the use of additional data analysis, that can further enhance the instrument sensitivity.

The first experiment on active cooling dates to 1953 by L.M.W. Milatz et al.\cite{Milatz1953}. Since then this techniques has been used in several field to enhance the sensitivity of the measurements apparatus. Some examples are the stabilization of the a kilogram-scale test mass in a gravity-wave interferometer\cite{Abbott2009}, the trapping of atomic\cite{Bushev2006} and sub-atomic particles\cite{DUrso2003} and the control of a mechanical oscillator at its thermal decoherence rate\cite{Wilson2015}. Among the various proposals\cite{Mertz, bar, ligo_feed} in the field of gravitational wave detectors, our work is certainly closer to the proposal of M. Pinard et al.\cite{Pinard} and D. Vitali et al.\cite{Vitali2001,Vitali1,Vitali2} where they suggest the use of a
non-stationary strategy to increase the detector sensitivity.

In this paper we report on experiments where, for the first time, it is demonstrated that opportunely acting on the cooling feedback it is possible to improve the signal-to-noise ratio during the recovery phase, when the apparatus is far from equilibrium.

\section{Experimental setup}
\begin{figure}
\includegraphics[width=\columnwidth]{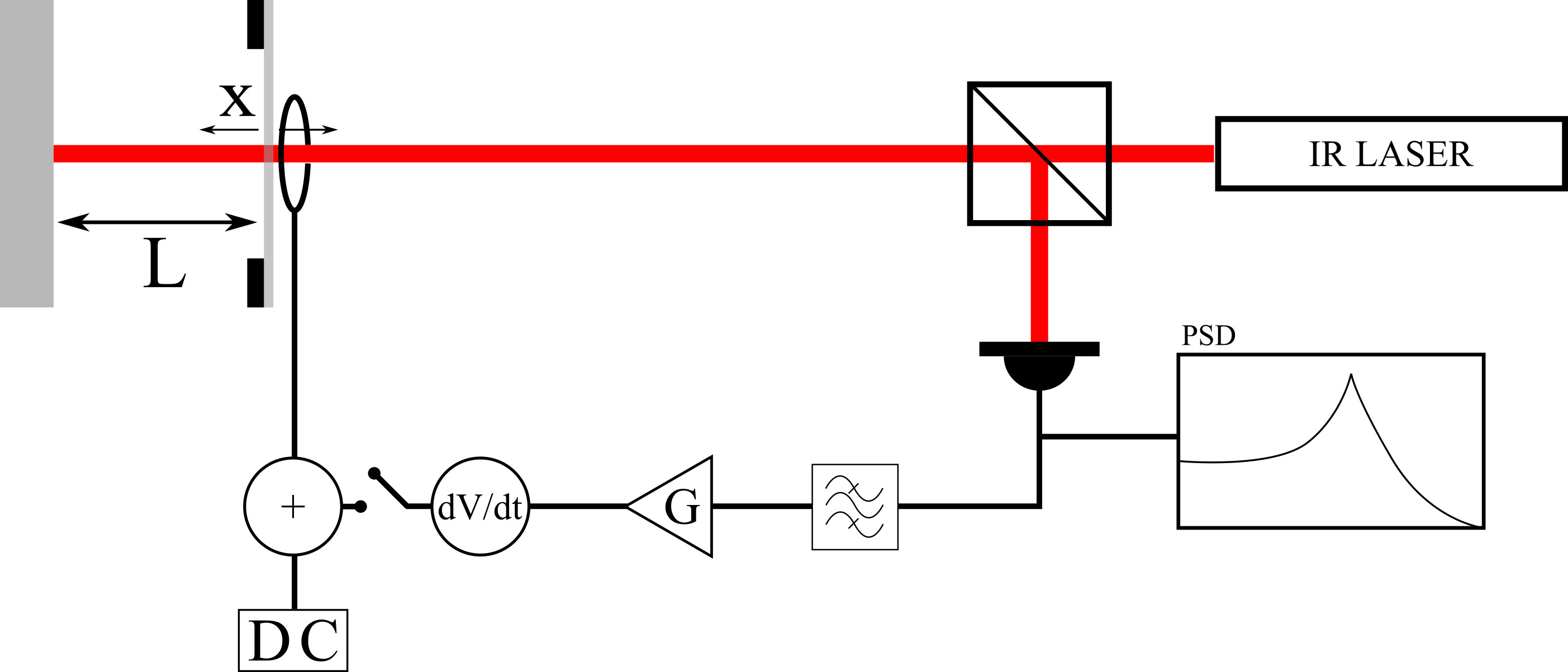}
\caption{\label{f:scheme_feedback}  Interferometer setup in single cavity configuration. A semi-reflective $Si_3N_4$ membrane is used as first mirror, part of the light is reflected from the membrane while the transmitted light is reflected from a second mirror placed at distance $L$ from the membrane. The two paths recombine at the photodetector creating interference. An electrostatic actuator is used to apply an electrostatic force on the membrane.}%
\end{figure}

In order to show the potential benefits of this approach, we performed an experiment by using a table top interferometer whose schematic  is depicted in Fig. \ref{f:scheme_feedback}. 
We set up a single cavity interferometer with two mirrors. The first mirror of the interferometer is composed by a semi-reflective $Si_3N_4$ membrane with a thickness of \SI{30}{\nano\meter} and a window size of \SI{5 x 5}{\milli\meter} produced by Norcada (NX10500XS). An infrared laser signal ($\lambda$ = \SI{1064}{\nano\meter}) is sent toward the membrane. Part of the light is reflected and part transmitted to the second mirror (BB1-E03 Thorlabs Broadband Dielectric Mirror) with a high coefficient of reflection ($>$\SI{99}{\percent}). The two beams recombine and the resulting interference is detected by a photodetector. The light intensity  on the photodetector can be expressed as:
\begin{equation}
I_p = I_0 \cos^2 \left( \frac{2 \pi}{\lambda} \left(L+x\right)\right)
\end{equation}
In this model we assumed that at the frequency of interest the thermal fluctuations of the second mirror are negligible since its effective mass, $M$ is much larger ($M =\SI{0.03}{\kilogram}$) than that of the membrane ($m = \SI{2.6e-9}{\kilogram}$). Therefore the only quantities of interest are the length of the cavity, $L$, and the position of the membrane, $x$.

Notice that, in the presented setup, the sensitivity of our interferometer is limited by the thermal noise of the membrane, and thus small variations of $L$ can be obfuscated by the stochastic motion of the membrane. This is a condition that mimics the response of a gravitational wave interferometer whose sensitivity to arm length variation is limited by noise affecting the test masses. Reducing the vibration of the membrane in the region of interest improves directly the sensitivity of the detector to $L$ variations. 

\begin{figure}
\includegraphics[width=\columnwidth]{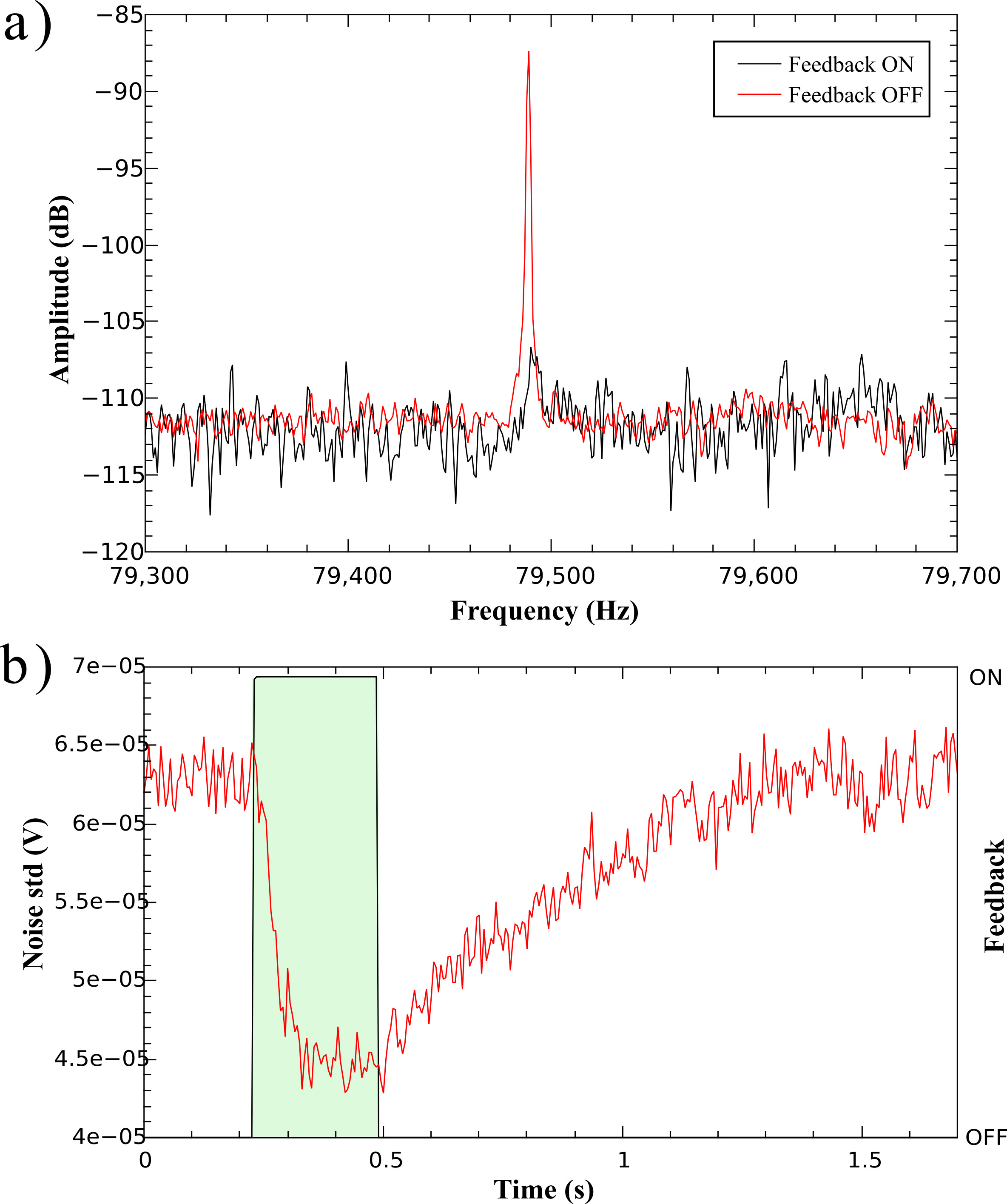}
\caption{\label{f:feedback} (a) Power spectral density of thermal noise of the cavity length setting the interferometer sensitivity with feedback deactivated and activated. (b) Time series of the effect of feedback force applied to the membrane. When the feedback is activated the energy of the noise around the first mode drops. Once the feedback is deactivated the thermal energy starts raising gathering energy from the thermal bath.}%
\end{figure}

\section{Feedback design}
In order to reduce the thermal noise induced oscillation of the membrane we have designed a feedback mechanism that acts on the membrane itself by means of an externally applied electrostatic force. The signal from the photodetector is filtered with a band pass filter with a width of \SI{300}{\hertz} around the resonance peak. The filtered signal is then amplified and differentiated. Finally a DC voltage is added to the signal in order to actuate correctly on the membrane. The schematic of the feedback mechanism is presented in Fig. \ref{f:scheme_feedback}. If we indicate with $x$ the position of the membrane, we can describe its dynamics as:
\begin{equation}
\ddot{x} = -\omega_0^2 x - \gamma_0 \dot{x} - \beta \dot{V} + \frac{1}{m}\xi(t)
\end{equation}
where $k=\omega_0^2 m$ the spring constant, $\gamma_0$ the damping constant, $\dot{V}$ the derivative of the band filtered voltage signal from the photodetector, $\beta$ the damping factor of the feedback and $\xi(t)$ the noise due to thermal fluctuations. When the feedback is active (not active) the value of $\beta>0$ ($\beta=0$). Considering that the voltage $V$ is proportional to the displacement $x$, the term $\beta\dot{V}$ acts as an additional damping mechanism on the membrane. While the presented model is for a linear system, the same effect of damping due to the feedback is expected for a non-linear system.
\begin{figure}
\includegraphics[width=\columnwidth]{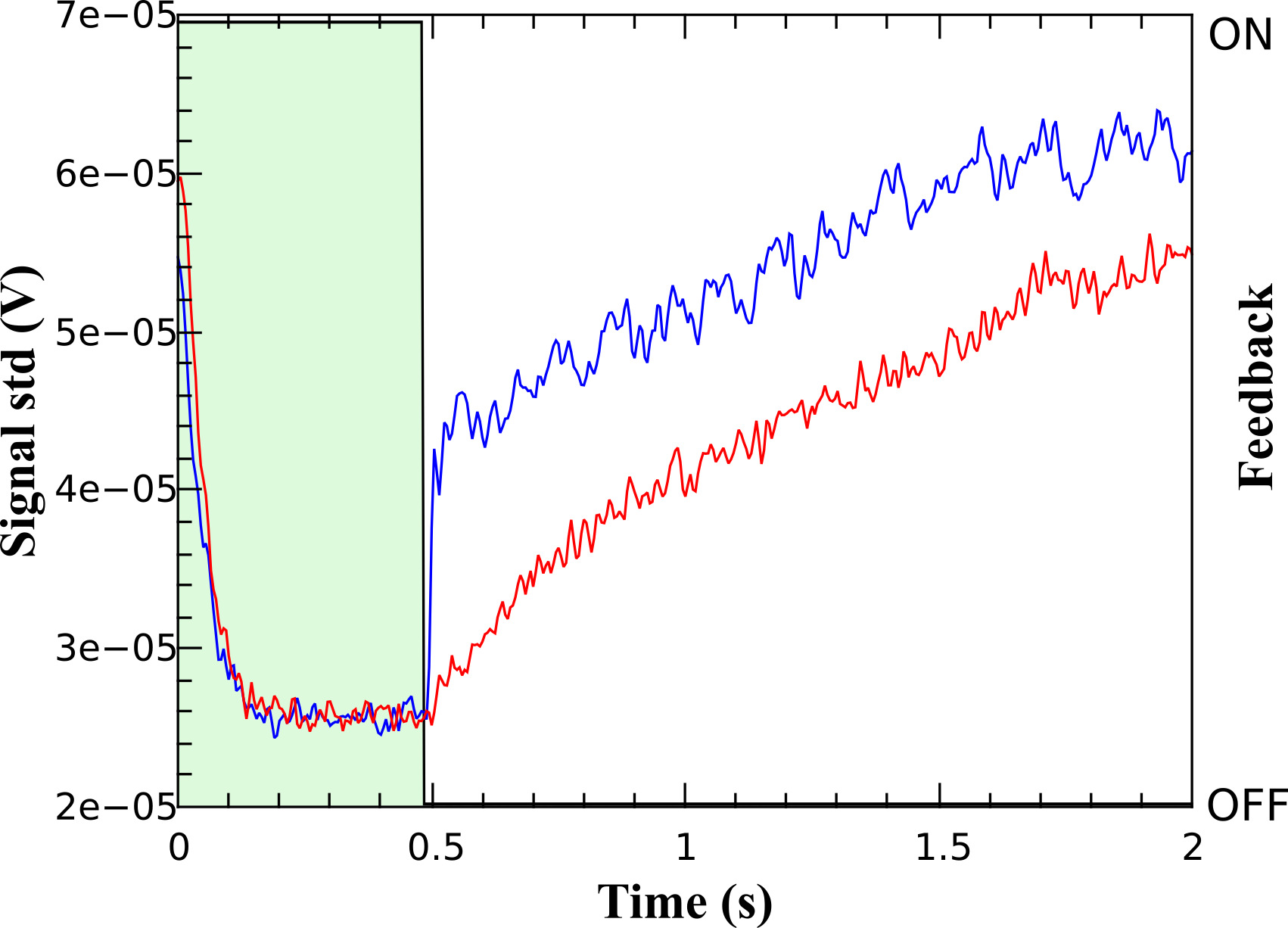}
\caption{\label{f:time-evolution} Time evolution of the detected signal std in the case where the periodic signal is absent (red curve) and when the periodic signal is present (blue curve). The periodic signal starts just after the feedback on the system is removed.}
\end{figure}

The power spectral density (PSD) of the voltage from the photodetector is presented in Fig. \ref{f:feedback}(a) in the case where the feedback is active and not active. The PSD in absence of a variation on $L$ sets the sensitivity curve of the detector. Notice that, in absence of the feedback, the thermal noise excites the membrane that responds mostly around its resonance frequency ($f_0$=\SI{79.485}{\kilo\hertz}), and thus a peak in the PSD is present. When the feedback is active the peak is reduced by \SI{20}{\dB}.

The evolution of the standard deviation (std) of the voltage at the photodetector, in a frequency range of $300$ Hz around the resonance peak, is presented in Fig. \ref{f:feedback}(b). Once the feedback is switched on (green region) the voltage std drops rapidly reaching its minimum floor set by the electronic noise of the measuring system. Once the feedback is switched off, restoring the original configuration, the membrane gathers energy from the thermal bath and thus the voltage std raises due to the raising of the membrane oscillation amplitude, approaching its thermal equilibrium value. 
The rate of increase of the voltage std is determined by the quality factor of the membrane (in our case $Q \sim 1.7e5$). The time evolution of the voltage std, once the feedback is switched on, can be expressed as an exponential time decay in the form of $A(t) = A \exp(-t/\tau_f)+c$, while the time evolution of of the voltage std, once the feedback is switched off, can be expressed as an exponential time growth in the form of $A(t) = A(1- \exp(-t/\tau_r))+c$. The resulting time constant for the falling and the rising are $\tau_f$ = \SI{0.045}{\second} and $\tau_r$ = \SI{0.707}{\second} respectively. It is interesting to notice that $\tau_f$ can be much smaller than $\tau_r$. This implies that a shorter time is required to cool down the system compared to the time required to relax to the thermal equilibrium. 

\section{Feedback performances}
To evaluate the impact of this technique on the interferometer sensitivity we compare the SNR relative to unprocessed data to the SNR achieved with active feedback. We have injected a periodic signal that produces a variation of the cavity length, $L$, by moving the second mirror at the same frequency of the membrane resonance peak. Such a signal mimics the presence of a gravitational wave target signal whose effect is to change the length $L$. The signal amplitude is chosen small enough (Signal-to-Noise Ratio (SNR) $\approx$ 0.3) to make it difficult to detect with standard interferometer detection schemes without use of feedback techniques or data processing.
We address two different cases: $i$) cooling in the absence of target signal; $ii$) cooling in the presence of the target signal.

\begin{figure}
\includegraphics[width=\columnwidth]{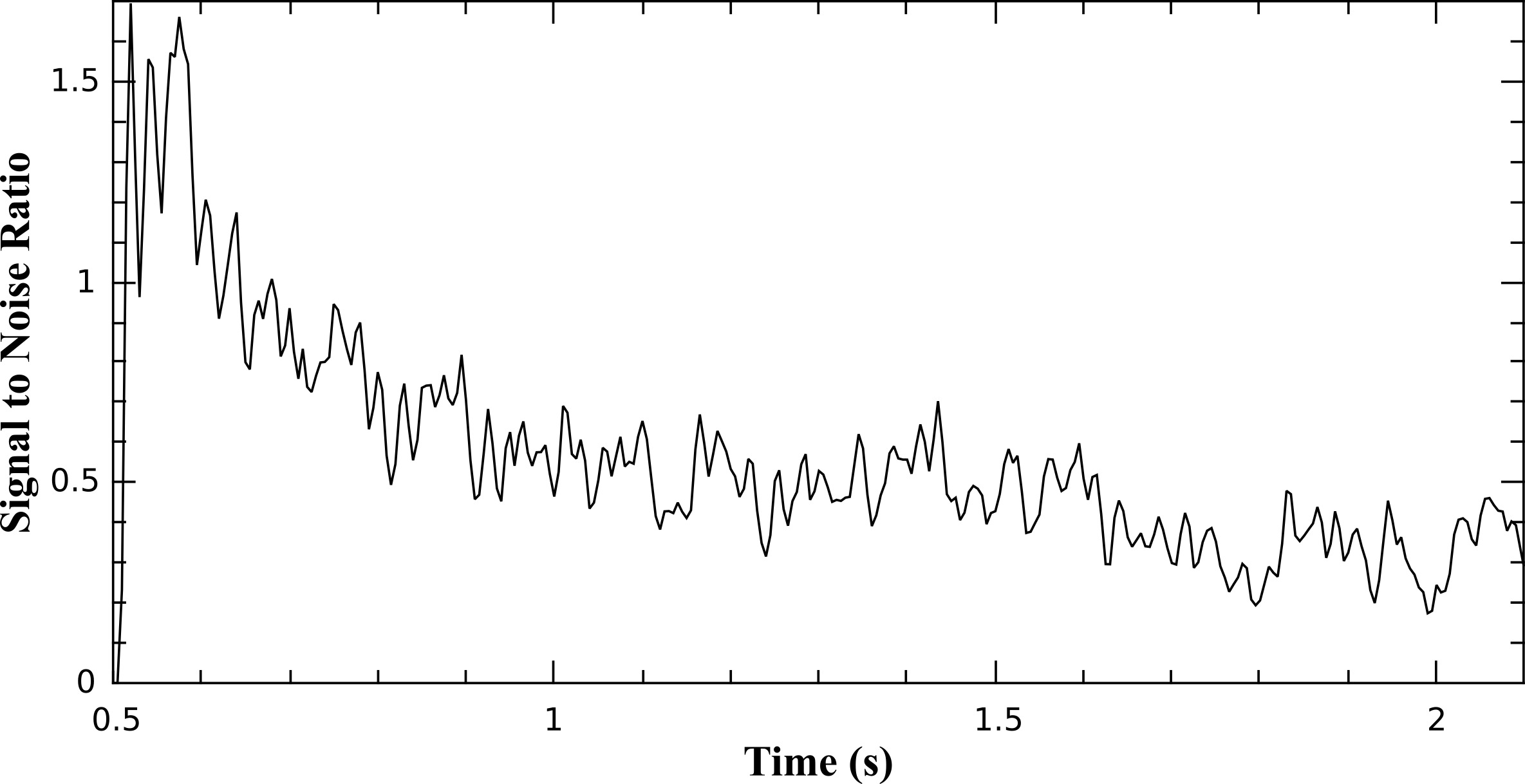}
\caption{\label{f:snr} SNR evaluated for a fixed amplitude signal as a function of time after the feedback is deactivated and the signal is present. The SNR is computed as the ratio between the power of the signal and the power of the noise, obtained from the data represented by the blue curve on Fig. \ref{f:time-evolution}. The SNR decreases as the membrane gather energy from the thermal bath reaching the final value when the membrane is at thermal equilibrium. This corresponds to the case where no active cooling is used.}
\end{figure}

$i$) At time $t=0$ s the feedback is turned on for $0.5$ s and the peak is cooled. At this time the injected signal is turned on and the voltage at the photodetector is monitored. In Fig. \ref{f:time-evolution} we present the voltage std in two different cases: with signal injected and without signal injected. While the power of the signal remains constant in time, the power of the random fluctuations increases as the system moves toward equilibrium with a rate determined by the time constant $\tau_r$. 
As it is well apparent, the overall std is higher when the injected signal is present, giving an initial value as large as SNR$ _{eq}\approx1.5$ compared to the original SNR$_{eq}\approx0.3$. Such a SNR improvement, however, tends to decrease as the system regain its thermal equilibrium (Fig. \ref{f:snr}).
Based on these results we can stress that in this case, operating the interferometer far from equilibrium, presents an advantage represented by a SNR that is significantly larger than that at equilibrium. Such an advantage SNR$-$SNR$_{eq}$ eventually decreases to zero when the system approaches thermal equilibrium, recovering its initial sensitivity.

$ii$) Let us suppose that the target signal exists at $t=-\infty$. As before, at $t=0$ s, the feedback is turned on for \SI{0.5}{\second}. The feedback signal acts here to reduce the variation of the cavity length $L$ and thus compensates both the thermal noise and the signal. The result is that the membrane is excited also with a signal equal to the signal to be detected but with phase inversion.
When the feedback is turned off, as before, the membrane will damp the oscillation at a rate defined by its quality factor. During this phase the SNR of the detector can be in principle worse than the original one (before the feedback) because it depends on the residual noise power at the feedback activation time: the lower is the power of the noise, the lower is the worsening of the SNR. The optimal case is obtained when the noise is reduced to zero and no worsening of the SNR is achieved (nor an increase). Thus in case $ii$) no improvement is expected by the application of the cooling strategy.

\begin{figure}
\includegraphics[width=\columnwidth]{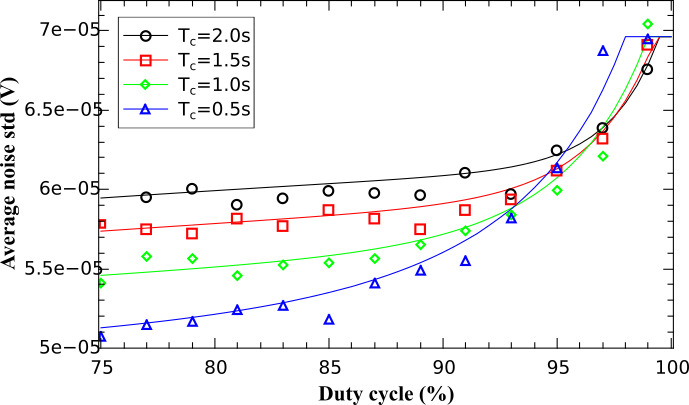}
\caption{\label{f:performances} Effect of feedback as function of duty cycle and total duration of the cycle, $T_c$. Dots represent the average standard deviation of the thermal noise around the resonance peak during the period in which the feedback is not active as function of the duty cycle of the feedback. Different colors represent different total window time. Dots are relative to experiment while continuous lines refer to data from the model.}%
\end{figure}

From this discussion it is evident that is of crucial importance to implement an effective feedback technique in order to minimize the noise power and thus maximize the SNR gain in case $i$) (signal not present during the cooling) and minimize the loss of detection performance in case $ii$) (signal present during the cooling).
Finally, we point out that during the application of the feedback the detector is not available for standard measuring in the frequency range of the feedback. It is thus important to minimize the time of feedback application maximizing the potential SNR enhancement. Notice that the latter condition can be expressed as the minimization of the average noise power during the period in which the feedback is active. Considering to operate the feedback cyclically we can define $t_f$ as the duration of feedback application and $T_c$ as the total duration of the cycle. We then define a duty cycle as $1-t_f/T_C$, the fraction of time where the detector is available. For a fixed $t_f$, increasing $T_c$ the duty cycle goes to one and the average noise power tends to the equilibrium value. Decreasing $T_c$ the duty cycle decreases and the average noise power decreases, reaching a minimum value when $T_c$ approach $t_f$. In order to evaluate the optimal choice of $T_c$ and $t_f$ we have performed several measurements varying these two parameters and monitoring the average noise std. The results are presented in Fig. \ref{f:performances}.
As expected, decreasing the duty cycle for a fixed cycle duration $T_c$, increases the performances of the system, i.e. the average noise std decreases. In general, for a fixed duty cycle it is more convenient to select a smaller cycle duration. However, this choice is bounded by two factors, one technical and one scientific. The technical factor depends on the technological implementation of the feedback, once activated the feedback requires a finite time to be operative (in our setup this corresponds to \SI{0.01}{\second}) limiting the minimum feedback time and thus the minimum cycle duration. This is clearly visible in Fig. \ref{f:performances} for $T_c$=\SI{0.5}{\second}, where we obtain a saturation for higher duty cycle values.
The scientific factor is associated with the duration of the signal to be detected. As commented above, if the feedback is activated while a signal is present, the SNR can be deteriorated, thus the cycle duration should be chosen in order to maximize the chances that a signal starts and ends inside a single cycle window. In particular this technique can be useful in conditions where the relaxation time of the resonators is much larger than the expected gravitational wave signal to be detected.

\section{Conclusions}
In conclusion we have  shown that selective cooling of single oscillatory modes of a detector mechanical structure is able to positively impact the measurement sensitivity of an interferometric detector by decreasing the fluctuation amplitude and increasing the SNR in the case of transient target signals. Experiments conducted with bench interferometers on thin silica membranes have shown promising results for the implementation of such technique in high sensitivity measurement apparatus that can be promising even in next generation gravitational wave detectors. It has to be underlined that this technique can be applied not only in the selective cooling of thermal noise peaks but also with other equilibrium fluctuation cases, like seismic or structural peaks in general.
Moreover we stress that the feedback cooling technique can be in principle applied also to frequency regions in the absence of any peak. In practice it easier to apply it to regions in the presence of resonances because there the amplitude of the oscillation is larger. One relevant example is represented by the violin modes of the Virgo and LIGO optics suspension wires\cite{virgo-violin,violin-ligo}. There, the low thermal noise suspension solution is based on the implementation of fused silica wires that connects to the test mass in order to form the so-called monolithic suspension\cite{4}. In this case the high quality factor of the fused silica wires will allow a most favourable duty cycle and, together with the novel sapphire suspension solution designed for the KAGRA interferometer\cite{Flaminio}, offer a potentially interesting test bench for this technology.

In general, our results suggest that, at difference with the standard way to operate gravitational wave detectors, i.e. under equilibrium conditions, the cyclical drive-and-relax to equilibrium operation might provide increased sensitivity in selected frequency ranges for the detection of transient gravitational signals.

\section*{Acknowledgments}
The authors gratefully acknowledge the financial support of the Istituto Nazionale di Fisica Nucleare (INFN) through the Virgo Project, the European Gravitational Observatory (EGO),  and the European Commission (FPVII, Grant agreement no: 318287, LANDAUER). The authors also gratefully acknowledge A. Giazotto  P. Amico, J. Kovalik, F. Travasso, F. Marchesoni and H. Vocca for useful discussions.

\end{document}